\documentclass[journal=jctcce,manuscript=article,layout=traditional]{achemso}
\usepackage{fullpage}
\usepackage{mathtools}
\usepackage{amsthm}
\usepackage{amssymb}
\usepackage{bm}
\usepackage[version=3]{mhchem}
\usepackage{braket}
\usepackage{relsize}
\usepackage{enumitem}
\usepackage[font=small,labelfont=bf]{caption}
\usepackage{graphicx}
\usepackage{footnote}
\usepackage{multirow}
\usepackage{tabularx}
\usepackage{booktabs}
\usepackage{makecell}
\usepackage{threeparttable}
\usepackage{chemfig}

\usepackage{caption}
\usepackage{subcaption}

\newcommand{\RNum}[1]{\expandafter{\romannumeral #1\relax}}

\newcolumntype{C}{>{\centering\arraybackslash}X}

\title{Efficient Calculation of NMR Shielding Constants Using Composite Method Approximations and Locally Dense Basis Sets
}	

\author{Jiashu Liang}
\altaffiliation{These authors contributed equally to this work.}
\affiliation{
	Kenneth S. Pitzer Center for Theoretical Chemistry,
	Department of Chemistry,
	University of California at Berkeley,
	Berkeley, CA 94720, USA
}
\author{Zhe Wang}
\altaffiliation{These authors contributed equally to this work.}
\affiliation{
	Kenneth S. Pitzer Center for Theoretical Chemistry,
	Department of Chemistry,
	University of California at Berkeley,
	Berkeley, CA 94720, USA
}
\author{Jie Li}
\affiliation{
	Kenneth S. Pitzer Center for Theoretical Chemistry,
	Department of Chemistry,
	University of California at Berkeley,
	Berkeley, CA 94720, USA
}
\author{Jonathan Wong}
\affiliation{
	Kenneth S. Pitzer Center for Theoretical Chemistry,
	Department of Chemistry,
	University of California at Berkeley,
	Berkeley, CA 94720, USA
}
\author{Xiao Liu}
\affiliation{
	Kenneth S. Pitzer Center for Theoretical Chemistry,
	Department of Chemistry,
	University of California at Berkeley,
	Berkeley, CA 94720, USA
}
\author{Brad Ganoe}
\affiliation{
	Kenneth S. Pitzer Center for Theoretical Chemistry,
	Department of Chemistry,
	University of California at Berkeley,
	Berkeley, CA 94720, USA
}
\author{Teresa Head-Gordon}
\affiliation{
	Kenneth S. Pitzer Center for Theoretical Chemistry,
	Department of Chemistry,
	University of California at Berkeley,
	Berkeley, CA 94720, USA
}
\author{Martin Head-Gordon}
\affiliation{
	Kenneth S. Pitzer Center for Theoretical Chemistry,
	Department of Chemistry,
	University of California at Berkeley,
	Berkeley, CA 94720, USA
}
\alsoaffiliation{
	Chemical Sciences Division,
	Lawrence Berkeley National Laboratory,
	Berkeley, CA 94720, USA
}
\email{mhg@cchem.berkeley.edu}

\date{\today}

\begin{document}

\begin{tocentry}
\includegraphics[width=6.2cm, height=3.5cm]{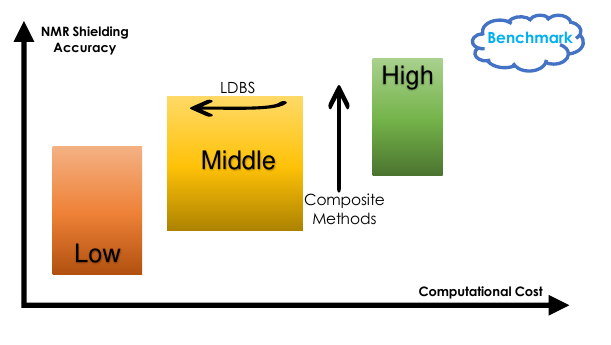}
\end{tocentry}

\begin{abstract}

This paper presents a systematic study of applying composite method approximations with locally dense basis sets (LDBS) to efficiently calculate NMR shielding constants in small and medium-sized molecules. The pcSseg-n series of basis sets are shown to have similar accuracy to the pcS-n series when n $\geq 1$ and can slightly reduce compute costs. We identify two different LDBS  partition schemes that perform very effectively for density functional calculations. We select a large subset of the recent NS372 database containing 290 H, C, N, and O shielding values evaluated by reference methods on 106 molecules to carefully assess methods of the high, medium, and low compute costs to make practical recommendations. Our assessment covers conventional electronic structure methods (DFT and wavefunction) with global basis calculations, as well as their use in one of the satisfactory LDBS approaches, and a range of composite approaches, also with and without LDBS. Altogether 99 methods are evaluated. On this basis, we recommend different methods to reach three different levels of accuracy and time requirements across the four nuclei considered. 

\end{abstract}

\maketitle

\clearpage

\section{Introduction} \label{sec:intro}

Nuclear Magnetic Resonance (NMR) spectroscopy is an indispensable tool for resolving molecular structures in organic chemistry and biochemistry research, especially when it is challenging to crystallize the target system and analyze it by X-ray crystallography.\cite{gil2011constitutional,becette2020solution, krivdin2019computationalH, krivdin2019computationalC} However, it is not always straightforward to map the molecular structure to the experimental spectra for complex systems. Therefore, \textit{ab initio} quantum chemistry now plays an increasingly important role in efforts to reduce ambiguities and confirm structures by predicting the spectrum as a function of stoichiometry or conformation. One of the primary observables determining NMR spectra is the magnetic shielding tensor, a second-order property defined at nucleus $A$, which can be defined as:
\begin{equation}
   \boldsymbol{\sigma}_A = \frac{\partial^2 E} { \partial \mathbf{M}_A \partial \mathbf{B}^{\mathrm{ext}} } 
                         = - \frac{\partial \mathbf{B}^{\mathrm{ind}}_A } { \partial \mathbf{B}^{\mathrm{ext}} }
\label{eq:shielding} 
\end{equation}
where $E$ is the molecular energy, $\mathbf{B}^{\mathrm{ext}}$ is the applied magnetic field, and $\mathbf{M}_A$ is the nuclear spin of $A$. The shielding thus determines the locally induced field, $\mathbf{B}^{\mathrm{ind}}_A = - \boldsymbol{\sigma}_A \mathbf{B}^{\mathrm{ext}} $. Eq. \ref{eq:shielding} indicates that $\boldsymbol{\sigma}_A$ is a (somewhat) spatially localized response to the externally applied field. Typically the isotropic shielding, $\sigma_A = \frac{1}{3}\mathrm{Tr} \boldsymbol{\sigma}_A$, is observed experimentally or simulated. Any electronic structure method can be used to approximate energy ($E$) and the shielding may then be evaluated as either an analytical or numerical derivative.

It is well-established that highly accurate methods such as the coupled-cluster theory with single and double excitations and perturbative triple excitations [CCSD(T)] together with large basis sets can provide reliable predictions of NMR shielding constants.\cite{teale2013benchmarking, schattenberg2021extended,gauss2002analytic,reid2015approximating}. These approaches are, however, impractical for any molecules with more than 10 non-hydrogen atoms due to their high computational cost. To study larger systems like proteins, which are often of current interest, various fragmentation methods have been developed.\cite{de1993methods, he2009protein, zhao2017accurate, kobayashi2018application,  herbert2019fantasy, chandy2020accurate} These methods employ the local property of NMR shielding, reducing the calculation time to linear scaling with molecule size without significant loss of accuracy (if the fragments are suitably chosen). However, suitable molecular fragments can sometimes contain more than 10 non-hydrogen atoms, which is still prohibitively expensive for high-accuracy calculations. Therefore, composite method approximations, a common tool of quantum chemistry for energy evaluations\cite{curtiss2011gn,narayanan2019accurate,thorpe2019high}, have been introduced to this area.\cite{kupka2011ccsd, kupka2013estimation, sun2013accurate, reid2015approximating, semenov2019calculation} Composite approaches employ different levels of theories and basis sets, usually combining high-level theory with a small basis set and low-level theory with a large basis set to approximate the results of high-level theory with a large basis set. Specifically, Reid et al. explored basic composites and double composites in detail and proved that some composite methods can accurately reproduce the CCSD(T)/large-basis-set results.\cite{reid2015approximating}

A key concern related to the accuracy-efficiency trade-off in NMR shielding calculations is the components of the basis set. Jensen found that different from energy calculations, some special basis functions like tight p-type functions can have a significant effect on predicting NMR shielding constants.\cite{jensen2008basis} Therefore, he designed a family of specialized basis sets, the pcS-n (n $=0-4$) sequence, which has been shown to converge NMR shielding constants faster than other basis set sequences.\cite{jensen2008basis, reid2014systematic, flaig2014benchmarking, jensen2016magnetic} Since the pcS-n sets are generally contracted, Jensen later developed segmented versions of pcS-n, which he called the pcSseg-n family.\cite{jensen2015segmented} With most quantum chemistry codes optimized for segmented basis sets, pcSseg-n calculations are expected to be faster than those with pcS-n series, with nearly equal accuracy. However, to the best of our knowledge, no paper has proven this from practical calculations.

Researchers can further utilize the locality of NMR shielding to optimize the computational costs associated with the size of the basis set. In the 1980s, Chesnut and Moore first introduced the idea of locally dense basis sets (LDBS), which assigns a large basis set only to the target atom (the dense part) and allocates smaller basis sets elsewhere in the molecule.\cite{chesnut1989locally} This can reduce the computation time substantially while keeping acceptable levels of accuracy. Their subsequent studies showed that one can obtain more accurate results when a multiatom segment or chemical functional group is selected as the dense part.\cite{chesnut1993use, chesnut1996use} Recently, Reid et al. performed a systematic study of partition schemes using Jensen’s pcS-n basis sets and recommended defining a dense group as a single non-hydrogen atom with connected hydrogens.\cite{reid2014systematic}

Since the composite method approximations and LDBS are designed for different aspects of NMR shielding calculations, it is possible to combine them to retain accuracy while gaining further saving on cost. In this work, we will study the accuracy-efficiency trade-off systematically by employing these two promising approximation methods and provide valuable references for researchers to choose the most suitable NMR calculation method based on their demands. First, we will briefly introduce the notations and computational details in the paper (Section~\ref{sec:method}). Then we will show the difference in accuracy and time requirement between pcS-n and pcSseg-n basis set series (Section~\ref{subsec:pcS}) and revisit the accuracy of different partition schemes of LDBS using pcSseg-n basis sets (Section~\ref{subsec:bench_LDBS}). Finally, we will provide recommendations regarding different accuracy requirements (Section~\ref{subsec:overall}). The conclusions are summarized in Section~\ref{sec:conclusion}.

\section{Methods} \label{sec:method}

\subsection{Composite method approximations}\label{subsec:composite}
A commonly used form in composite method approximations is to approximate a computationally expensive target model, $T_{\text{high}}/B_{\text{large}}$, using 3 computationally much cheaper calculations:
\begin{equation}
    \begin{aligned}
    T_{\text{high}}/B_{\text{large}} &\approx T_{\text{low}}/B_{\text{large}} + \left( T_{\text{high}}/B_{\text{small}} - T_{\text{low}}/B_{\text{small}} \right) \\
    &= T_{\text{high}}/B_{\text{small}} 
    + \left( T_{\text{low}}/B_{\text{large}} - T_{\text{low}}/B_{\text{small}} \right)
    \end{aligned}
\label{eq:composite}
\end{equation}
Here $T_{\text{high}}$ and $T_{\text{low}}$ are two levels of theory and $B_{\text{large}}$ and $B_{\text{small}}$ are two basis sets with different sizes. The composite energy defined by Eq. \ref{eq:composite} can then be used to evaluate chemical shielding constants via Eq. \ref{eq:shielding}.

This model can be viewed as correcting a low level of theory in a large basis set for missing correlation effects (first line of Eq. \ref{eq:composite}) on the assumption that such effects can be captured in a small basis set. Or, it can be equivalently viewed as correcting a high level of theory in a small basis set for missing basis set effects on the assumption that such effects can be captured at a lower level of theory (second line of Eq. \ref{eq:composite}). Either view can be justified based on perturbation theory arguments, although similar rates of convergence of the energy with basis set for $T_{\text{high}}$ and $T_{\text{low}}$ are desirable. In other words, the composite approach implicitly assumes that the incomplete basis set error of chemical shielding constants at $T_{\text{high}}$ and $T_{\text{low}}$ levels are of comparable size, which is usually true for density functional theory (DFT), second-order M{\o}ller{\textendash}Plesset perturbation theory (MP2), and CCSD(T) methods.

In this paper, we denote a composite method as $T_{\text{high}}\left( B_{\text{small}} \right) \cup T_{\text{low}} \left( B_{\text{large}} \right)$. We choose pcSseg-3 (or pcSseg-3 for the dense region in LDBS) as $B_{\text{large}}$, and we select pcSseg-1 as $B_{\text{small}}$. On the theory side, two levels of accuracy will be investigated: 1) MP2 or double hybrid (DH) DFT as $T_{\text{high}}$ and lower rungs\cite{perdew2005prescription} of DFT as $T_{\text{low}}$; 2) CCSD(T) as $T_{\text{high}}$ and MP2 or DHDFT as $T_{\text{low}}$.

\subsection{Locally dense basis set}\label{subsec:LDBS}

The local nature of NMR shielding tensors motivates the idea of LDBS as an approach to facilitate calculations. In this work, we explore two kinds of partition schemes. The first one is based on Reid et. al's recommendations.\cite{reid2014systematic} We regard a target non-hydrogen atom and its bonded hydrogen atoms as a group and denote it as pcSseg-XYZ. X refers to the allocated (large) basis set of the target group. Similarly, Y refers to the chosen basis set of nearest-neighbor groups, while Z refers to the (smallest) basis set used for more distant groups. We choose pcSseg-321 and pcSseg-331 here. For a detailed comparison, please consult Ref~\citenum{reid2014systematic}. The other scheme selects chemical functional groups according to Chesnut’s suggestions.\cite{chesnut1993use} We denote an LDBS using this approach as pcSseg-func-XYZ, and we will choose pcSseg-func-321 as an example to explore. More details on the implementation of the partitioning process are described in the first section of the Supporting Information.

\subsection{Computational details}\label{subsec:comp}

Four sets of molecules are chosen as our data sets for different purposes:

\begin{enumerate}
    \item NS372 set: Shielding constants at H, C, N, and O nuclei of the large NS372 set\cite{schattenberg2021extended} are chosen as our overall benchmark reference in Section~\ref{subsec:bench_LDBS} and Section~\ref{subsec:overall}, which provides a quite comprehensive assessment of the light main-group elements with CCSD(T)/pcSseg-3 reference data. This set comprises 290 shielding values of 106 molecules containing 123 $^{1}$H, 93 $^{13}$C, 43 $^{15}$N, and 31 $^{17}$O after discarding BH for its large static correlation. The molecular geometries are directly adopted from the Supporting Information of the NS372 paper.\cite{schattenberg2021extended}
    \item NS212 set: A subset of NS372 containing 89 molecules and 212 nuclei evaluated at the CCSD(T)/pcSseg-4 level is used for comparison of accuracy and efficiency of the pcS-n and pcSseg-n series in Section~\ref{subsec:pcS}.
    \item M20 set: a set of twenty larger molecules with various common functional groups is applied to assess the effect of different LDBS partition schemes in Section~\ref{subsec:bench_LDBS}. M20 is needed because the smaller molecules in the NS372 set may not contain enough ``environment'' to permit meaningful assessment of different LDBS partition schemes. Q-Chem 5.4 software\cite{epifanovsky2021software} is used to optimize the molecule structures at the $\omega$B97X-V/aug-cc-pVTZ level\cite{kendall1992thom} after MMFF94 force field\cite{halgren1996merck} pre-optimization.
    \item Time evaluation set: three molecules containing two non-hydrogen atoms, three molecules containing four non-hydrogen atoms, and three molecules containing eight non-hydrogen atoms are collected to test the time cost of different methods.
\end{enumerate}

The CCSD(T) method was used to generate the reference chemical shielding constants in the NS372 data set,\cite{schattenberg2021extended} because it should be reliable for molecules without strong static correlation. \cite{teale2013benchmarking, schattenberg2021extended,gauss2002analytic,reid2015approximating} For example, a comparison of nuclear magnetic shielding constants for HF, CO, N$_2$, and N$_2$O with the qz2p basis set shows a difference of 0.01 ppm for $^1$H, 0.5 ppm for $^{13}$C, and around 1 ppm for $^{15}$N and $^{17}$O between CCSD(T) and the higher level CCSDT method. \cite{gauss2002analytic} Another study on methanol shows a deviation of 0.001-0.006 ppm for $^1$H, 0.14 ppm for $^{13}$C, and 0.28 ppm for $^{17}$O. \cite{auer2009high} The difference increases to dozens of ppm for molecules with large static correlation (for example, $^{17}$O in O$_3$, which is, however, not included in our benchmark set). Compared to experimental values, the estimated mean absolute error (MAE) of CCSD(T) at the complete basis set limit (CBS) is also small, on the order of 0.15 ppm for hydrogen nuclei, 0.4 ppm for carbon, 3 ppm for nitrogen, and 4 ppm for oxygen. \cite{teale2013benchmarking, reid2015approximating}. The accuracy of the pcSseg-3 basis set was also evaluated by the developers of the NS372 dataset. The average changes range from 0.2 to 0.4 ppm for second-row nuclei and 0.03 ppm for hydrogen nuclei when the basis set is switched from pcSseg-3 to pcSseg-4, \cite{schattenberg2021extended} which is acceptable. Based on the data shown here, we decided that our target error for this work is below 0.1 ppm for H nuclei, 1 ppm for C, 3 ppm for N, and 4 ppm for O.
 
We performed CCSD(T) shielding calculations with the CFOUR program package, version 2.1.\cite{matthews2020coupled,stanton2010cfour,harding2008parallel} All other calculations, if not specified, were carried out using ORCA 5.0.3.\cite{neese2020orca} For all calculations carried out with ORCA, self-consistent field (SCF) convergence was set to $10^{-9}$ while the coupled perturbed self-consistent field convergence was set to a threshold of $10^{-7}$. For DFT calculations, local xc integrals were calculated over ORCA default grid DefGrid3 for all atoms, which is accurate enough for our purposes. Gauge-including atomic orbitals (GIAOs) were employed in all calculations. The resolution of identity approximation (RI) was used for double hybrid DFT (DHDFT) and most MP2 calculations, with the cw5C\cite{hattig2005optimization} auxiliary basis set. For further acceleration,\cite{stoychev2018efficient} def2-JK\cite{weigend2008hartree} auxiliary basis set was employed for the Coulomb and exchange part of MP2 in Sections~\ref{subsec:pcS} and \ref{subsec:bench_LDBS}. The pcSseg-n basis sets are used in Section~\ref{subsec:bench_LDBS} and Section~\ref{subsec:overall} following the conclusions of Section~\ref{subsec:pcS}. Basis sets not built in the computational packages were downloaded from Basis Set Exchange (http://www.basissetexchange.org/).\cite{pritchard2019new}

The performance of DFT functionals for predicting magnetic shielding has been extensively benchmarked in previous work.\cite{flaig2014benchmarking, schattenberg2021extended, de2021double}. We selected the following functionals from each rung based on their reported performance and popularity for our present assessment: B97-D (Rung 2),\cite{grimme2006semiempirical} KT3 (Rung 2),\cite{keal2004semiempirical} B97M-V (Rung 3),\cite{mardirossian2015mapping} SCAN (Rung 3),\cite{sun2015strongly} M06-L (Rung 3),\cite{zhao2006new} PBE0 (Rung 4),\cite{adamo1998toward,adamo1999toward} $\omega$B97X-V (Rung 4),\cite{mardirossian2014omegab97x} $\omega$B97X-D3 (Rung 4),\cite{lin2013long} B2GP-PLYP (Rung 5),\cite{karton2008highly} and DSD-PBEP86 (Rung 5).\cite{kozuch2011dsd,kozuch2013spin} KT3 is provided by LibXC\cite{lehtola2018recent} within the ORCA framework. For these $\tau$-dependent meta-GGAs, Dobson’s $\tau_{\text{D}}$ model\cite{dobson1993alternative} is used for B97M-V while the ORCA default $\tau_{\text{GI}}$ model\cite{schattenberg2021effect} is used for others in the light of their reported performance.\cite{schattenberg2021extended} It is worth mentioning that the $\tau_{GI}$ model is not physically well-justified and its good performance is possibly due to error cancellation.

All timing jobs were run on a single Haswell node of the NERSC supercomputer. Each Haswell node (Intel Xeon Processor E5-2698 v3) has two sockets, each populated with a 2.3 GHz 16-core Haswell processor. The computational cost is evaluated by averaging the wall time of single computation tasks of molecules with the same number of non-hydrogen atoms across the time evaluation set.

\section{Results and Discussions} \label{sec:bench_method}

\subsection{Comparison of accuracy and efficiency of pcS-n and pcSseg-n series}\label{subsec:pcS}

\begin{figure}[ht!]
    \centering
    \includegraphics[width=\textwidth]{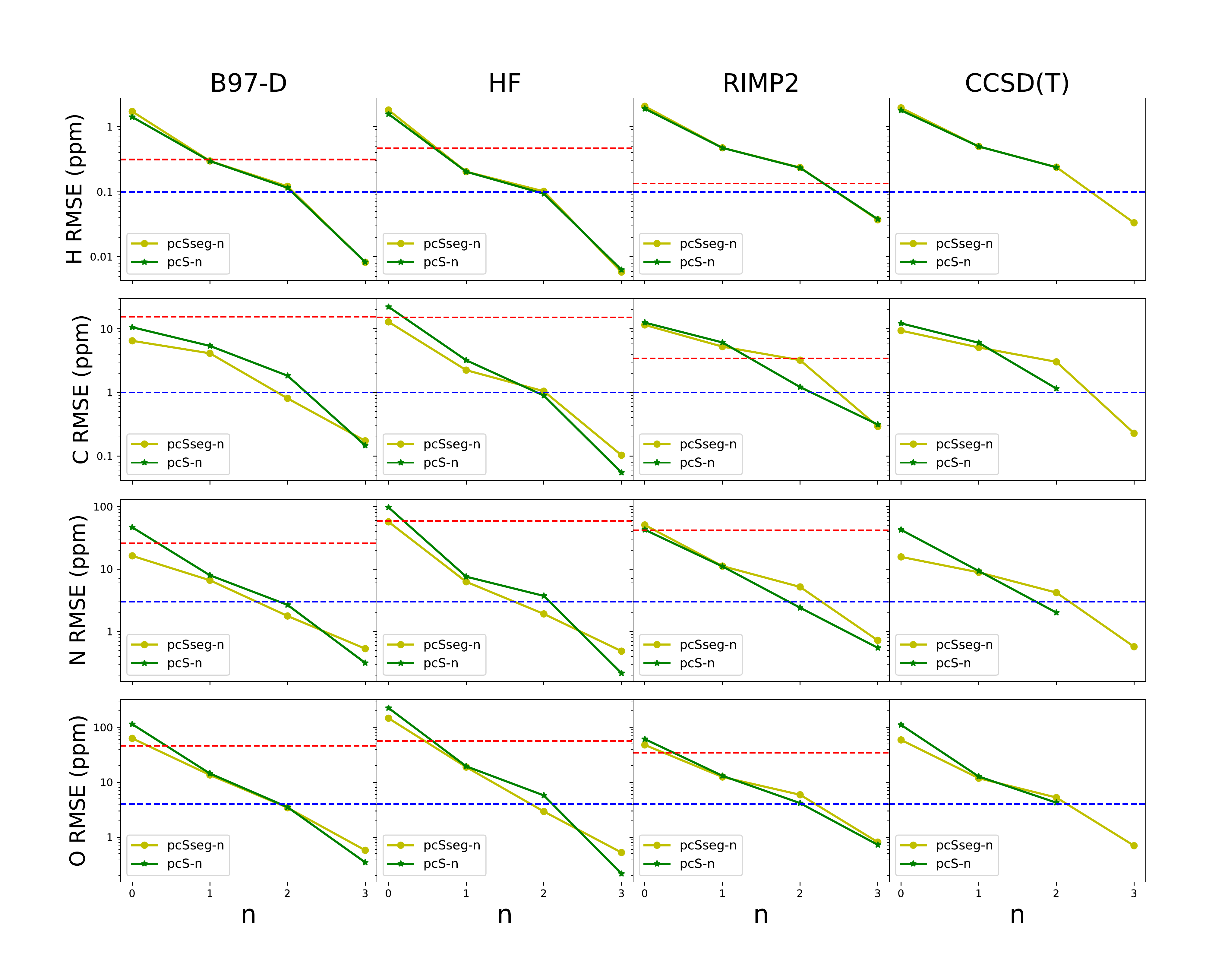}
    \caption{Comparison of the RMSEs (in ppm) of pcS-n and pcSseg-n series (n $\leq 3$) for B97-D, HF, RIMP2, and CCSD(T) shielding calculations with respect to pcSseg-4 results with the same method on H, C, O, and N nuclei. The red horizontal lines indicate the intrinsic method error, represented by the RMSEs of the method with pcSseg-4 against CCSD(T)/pcSseg-4. The blue horizontal lines indicate the target error: 0.1 ppm for H nuclei, 1 ppm for C, 3 ppm for 
    N, and 4 ppm for O.}
    \label{fig:basis}
\end{figure}

We first explore the basis set convergence of magnetic shieldings using the pcS-n and pcSseg-n series for DFT (taking B97-D as the representative functional) and wavefunction theory [i.e., HF, resolution of identity MP2 (RIMP2), and CCSD(T)]. Figure~\ref{fig:basis} displays the Root-Mean Square Errors (RMSEs) of H, C, O, and N nuclei as a function of n compared to the same method using a CBS (approximated with pcSseg-4 basis set here) on NS212 set.

Comparing the panels of Figure~\ref{fig:basis} horizontally reveals that DFT and HF converge faster than post-HF methods [RIMP2 and CCSD(T)] for both pcS-n and pcSseg-n series, consistent with earlier results\cite{reid2014systematic}. Slower convergence of post-SCF methods reflects the polynomial convergence of the wavefunction-based correlation energy with the cardinal number of the AO basis. If we consider the intrinsic method errors, double-zeta basis sets (pcS-1 or pcSseg-1) are sufficient for the use of DFT and HF while triple- or even quadruple-zeta basis sets (n $ \geq 2$) are required to achieve the best performance of post-HF methods. Fortunately, the basis set errors of these different methods for shieldings are still of the same magnitude, suggesting that composite correction methods could be successful. 

When examining Figure~\ref{fig:basis} more closely, we observe that the pcS-n and pcSseg-n series behave similarly on H nuclei, but their RMSEs on non-hydrogen nuclei cross---pcSseg-0 outperforms pcS-0 for nearly all methods and pcSseg-2 outperforms pcS-2 for DFT and HF, while pcS-2 outperforms pcSseg-2 for post-HF methods and pcS-3 outperforms pcSseg-3 for DFT and HF. However, the differences are quite small compared to the method error (also illustrated in Figure~S4). The elapsed (wall) times for calculations with the pcSseg-n and pcS-n series are similarly quite close. Table~\ref{tab:basis_time} shows the difference is only around 5-20\% except for the CFOUR program. This indicates that either series can be applied in practice. We will utilize pcSseg-n basis sets in the following subsections because the basis sets employed in this study are primarily double-zeta and quadruple-zeta, where the pcSseg-n series takes a bit less time.

\begin{table}[ht!]
\caption{Comparison of average wall time (in hours) of pcS-n and pcSseg-n basis sets using different methods and different programs for molecules with 8 non-hydrogen atoms. Only one physical core is used here.}
\begin{tabular}{lcccc}
\hline
Basis set & \multicolumn{1}{l}{B97-D (ORCA)} & \multicolumn{1}{l}{B97-D (Q-Chem)} & \multicolumn{1}{l}{RIMP2 (ORCA)} & \multicolumn{1}{l}{MP2 (CFOUR)} \\ \hline
pcS-0     & 0.015                            & 0.020                              & 0.158                             & 0.023                             \\ 
pcSseg-0  & 0.016                            & 0.022                              & 0.156                             & 0.025                             \\ \hline
pcS-1     & 0.059                            & 0.050                              & 0.286                             & 0.358                             \\
pcSseg-1  & 0.054                            & 0.048                              & 0.276                             & 0.366                             \\ \hline
pcS-2     & 0.55                             & 0.45                               & 1.49                              & 7.32                              \\ 
pcSseg-2  & 0.50                             & 0.40                               & 1.57                              & 13.93                             \\ \hline
pcS-3     & 6.45                             & 10.51                              & 18.65                             &                                   \\ 
pcSseg-3  & 5.69                             & 10.01                              & 16.59                             &                                   \\ \hline
\end{tabular}
\label{tab:basis_time}
\end{table}

\subsection{Relative accuracy and computational cost of 3 different LDBS partition schemes.}\label{subsec:bench_LDBS}

We chose to assess three different partition schemes, labeled as pcSseg-321, pcSseg-331, and pcSseg-func-321, in the notation defined in Section~\ref{subsec:LDBS}. These LDBS partition schemes are compared for DFT (B97-D as the representative) and wavefunction theory (RIMP2 as the representative).

\begin{figure}[ht!]
    \centering
    \includegraphics[width=0.95\textwidth]{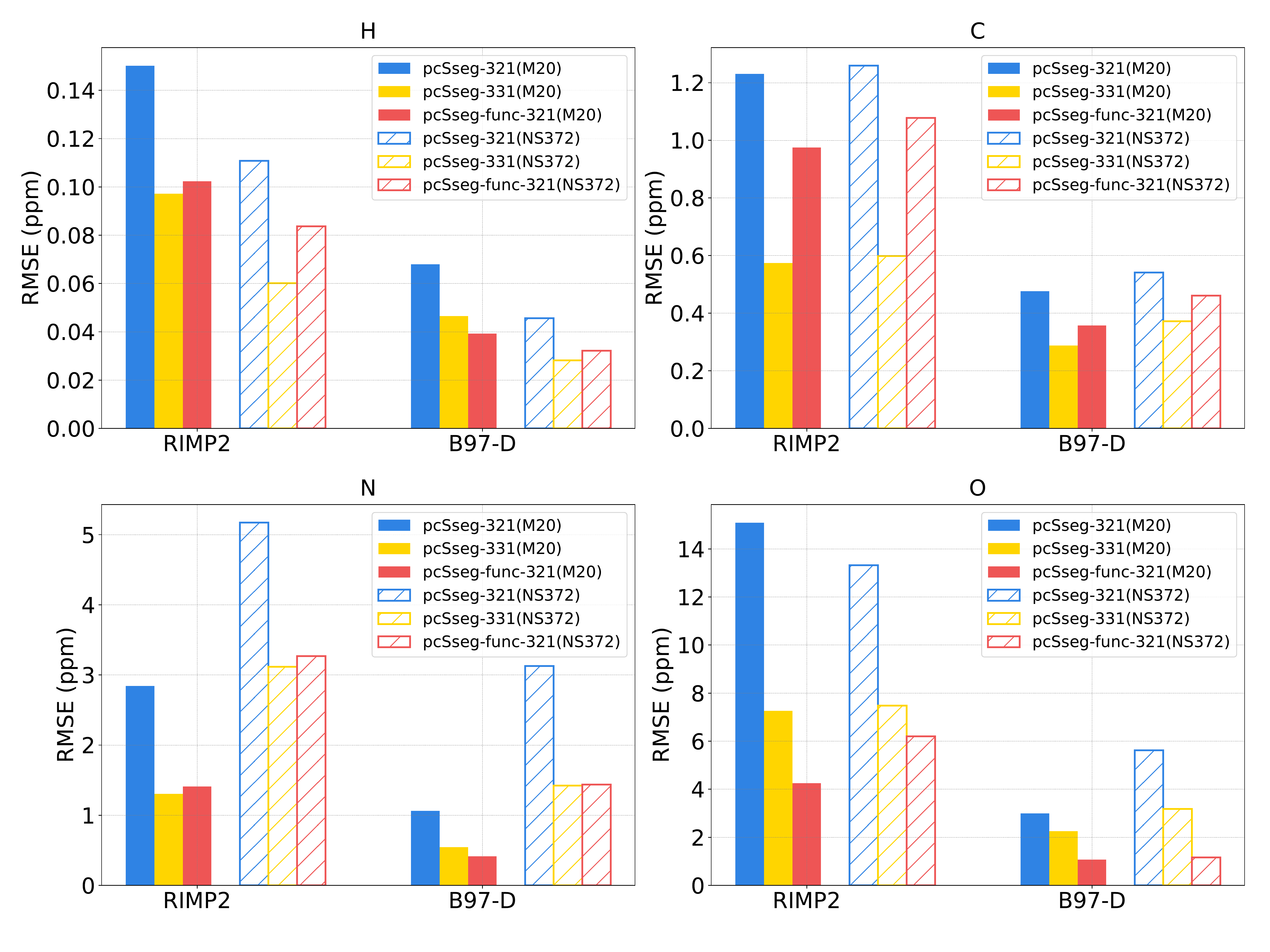}
    \caption{Comparison of the RMSEs (in ppm) of pcSseg-321, pcSseg-331, and pcSseg-func-321 for RIMP2 and B97-D with respect to each method itself with pcSseg-3 for NMR shieldings on H, C, O, and N nuclei contained in two different benchmark sets, represented by solid bars and dashed line bars respectively.}
    \label{fig:ldbs_vsself}
\end{figure}

Figure~\ref{fig:ldbs_vsself} shows the RMSEs with the 3 different LDBS partition schemes for RIMP2 and B97-D taking their CBS value (approximated with the pcSseg-3 basis set) as the reference on the M20 and NS372 data set. Although the composition of molecules in the two data set are different, the findings drawn from the two data sets, however, remain the same. First, consistent with the global basis set convergence trends seen in Section~\ref{subsec:pcS}, we find that the error induced by the LDBS approximation is much lower for B97-D than RIMP2 on all four elements. Shielding constants calculated at the RIMP2 level are more sensitive to the choice of basis set, implying that LDBS may work better for DFT and DFT-based composite methods. Second, pcSseg-321 consistently performs the worst. The RMSE of pcSseg-331 and pcSseg-func-321 are comparable for H and N nuclei, whereas pcSseg-func-321 prevails for the O nucleus and pcSseg-331 prevails for the C nucleus. It is worth noting that pcSseg-func-321 does not recognize any functional groups with more than 4 non-hydrogen atoms and thus uses fewer basis functions than pcSseg-331 when describing some important chemical structures like aromatic rings. Therefore, the trends in Figure~\ref{fig:ldbs_vsself} tend to reflect the element-specific numbers of basis functions included in pcSseg-331 versus pcSseg-func-321 versus pcSseg-321. 

\begin{figure}[ht!]
    \centering
    \includegraphics[width=0.95\textwidth]{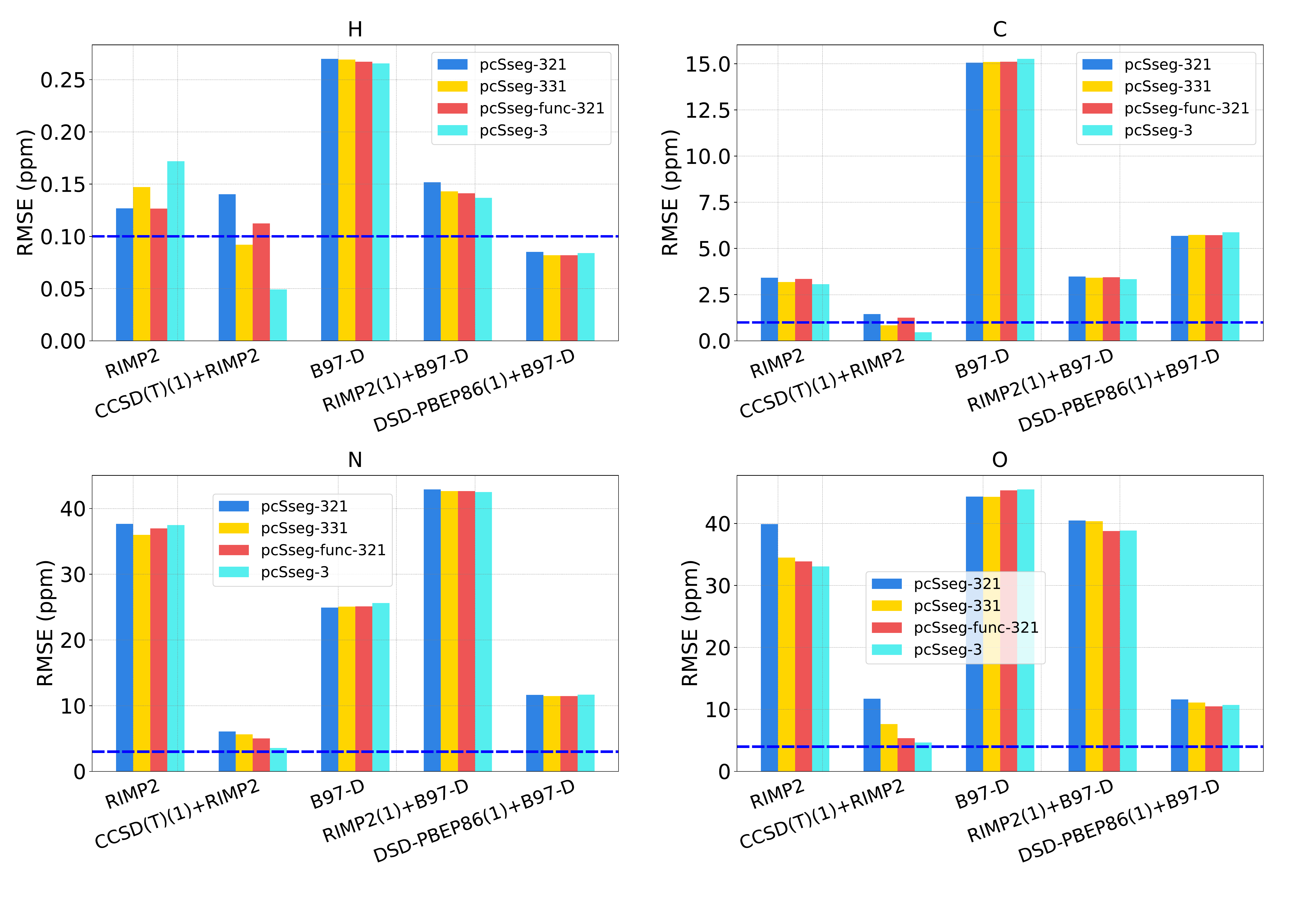}
    \caption{Comparison of the RMSEs (in ppm) of pcSseg-321, pcSseg-331, pcSseg-func-321, and pcSseg-3 for RIMP2, B97-D, and related composite methods with respect to CCSD(T)/pcSseg-3 on H, C, O, and N nuclei of the NS372 set. Different basis sets to be compared work as larger basis sets ($B_{\text{large}}$) in composite methods. The blue horizontal lines indicate the target error: 0.1 ppm for H nuclei, 1 ppm for C, 3 ppm for N, and 4 ppm for O.}
    \label{fig:ldbs_vsCCSDt}
\end{figure}

Additionally, RMSEs for RIMP2, B97-D, and certain composite methods using various LDBS partition schemes with regard to CCSD(T)/pcSseg-3 on the NS372 set are displayed in Figure~\ref{fig:ldbs_vsCCSDt}. It is evident that the use of LDBS has little to no impact on the RMSE for B97-D and related composite approaches. However, the error associated with the LDBS is more noticeable for the highly accurate composite method [i.e., CCSD(T) as the high-level theory and RIMP2 as the low-level theory], where the RMSE increases above the target error.

Table~\ref{tab:ldbs_time} compares the computational cost for RIMP2 and B97-D with various LDBS partition schemes and the global pcSseg-3 basis set on molecules with 8 non-hydrogen atoms. The LDBS technique (pcSseg-func-321 and pcSseg-321) can save more than half of the computational time and the reduction is expected to grow for larger molecules. For a single molecule, we need to calculate different numbers of jobs under different partition schemes. Usually, pcSseg-func-321 will have fewer jobs than pcSseg-331 and pcSseg-321. Therefore, we believe that pcSseg-func-321 is the best partition scheme assessed in terms of accuracy and computing efficiency and we employ it in Section~\ref{subsec:overall}.

\begin{table}[ht!]
    \caption{Comparison of average wall time (in hours) for three molecules with 8 non-hydrogen atoms using the pcSseg-321, pcSseg-331, pcSseg-func-321, and pcSseg-3 basis sets for RIMP2 and B97-D. The LDBS wall times are calculated by summing all component jobs needed. We use MPI parallelization with four physical cores here, and shieldings are evaluated at all nuclei.}
    \centering
        \begin{tabular}{l c c c c}
        \hline
        & Average basis functions & $N_{\text{total jobs}}$ & RIMP2 & B97-D \\
        \hline
        pcSseg-321 & 336 & 24 & 1.977 & 0.635 \\
        pcSseg-func-321 & 354 & 21 & 1.994 & 0.645 \\
        pcSseg-331 & 448 & 18 & 3.979 & 1.367 \\
        pcSseg-3 & 917 & 3 & 4.390 & 1.554 \\
        \hline
        \end{tabular}
        \label{tab:ldbs_time}
\end{table}

As shown in Table \ref{tab:ldbs_time}, the compute advantage of the LDBS approach is already useful even when evaluating NMR shieldings at all nuclei in a medium-sized molecule. Larger speedups can be obtained in some special cases. An interesting  example is when shieldings are only needed at a single nucleus (or within a single functional group in the LDBS pcSseg-func-321 approach). The speedup then approaches the ratio of pcSseg-3 time to the pcSseg-1 time ($\sim 4^3 - 4^4$). Another scenario in which that same speedup is approached is when using double numerical differentiation of energies with finite applied fields and nuclear spins to obtain the shielding. 

\subsection{Overall benchmark}\label{subsec:overall}

\begin{figure}
    \centering
    \includegraphics[width=0.7\textwidth]{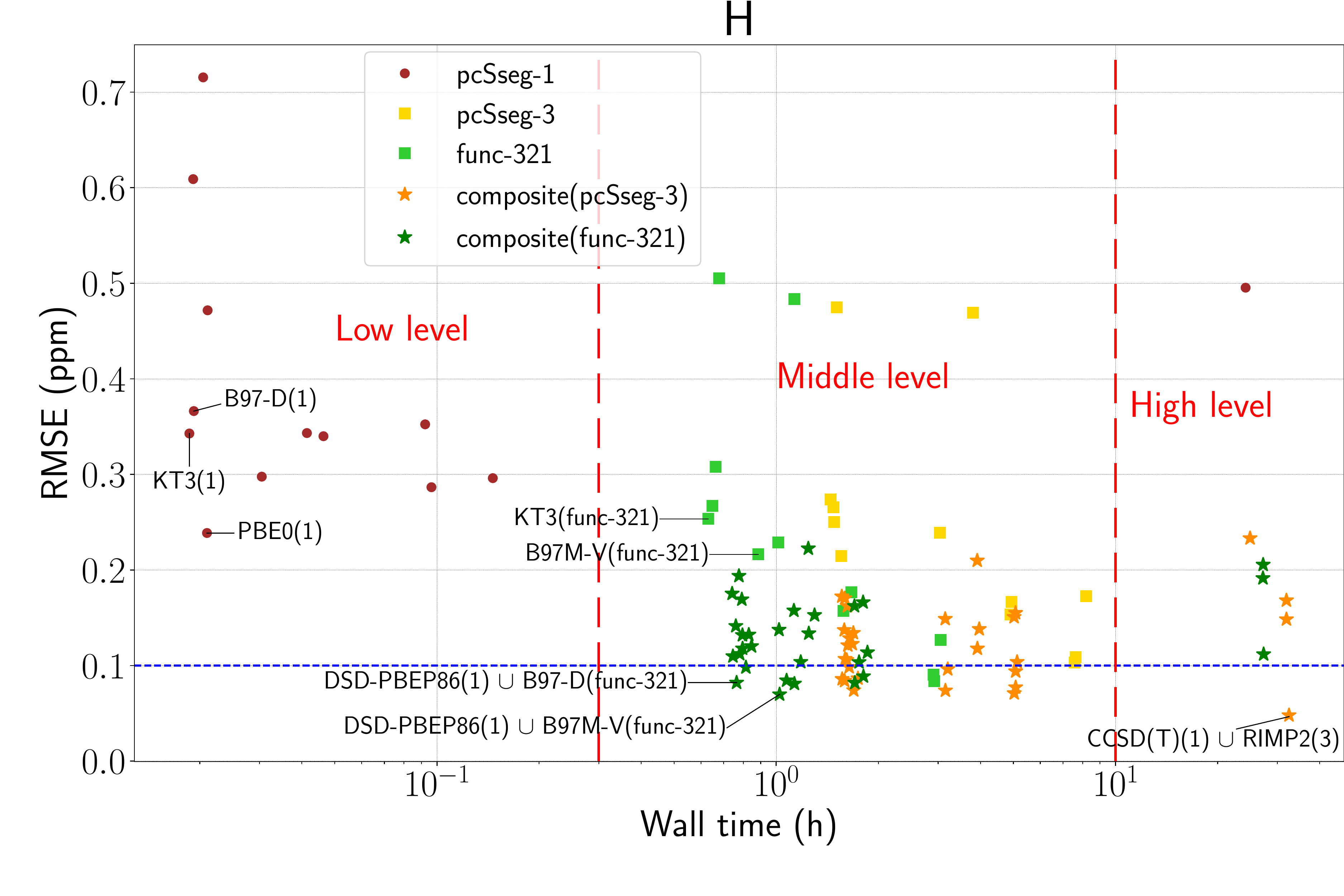}
    \caption{Comparison of the RMSEs for proton shieldings (in ppm) of different methods with different basis sets relative to CCSD(T)/pcSseg-3 on the NS372 set against the average wall time (in hours) for molecules with eight non-hydrogen atoms calculated with four physical cores. ``(X)" in the method labels represents the basis set pcSseg-X. Only the recommended methods are labeled here. The blue horizontal lines indicate the target error: 0.1 ppm for H nuclei.}
    \label{fig:overall_H}
\end{figure}
\begin{figure}
    \centering
    \includegraphics[width=0.7\textwidth]{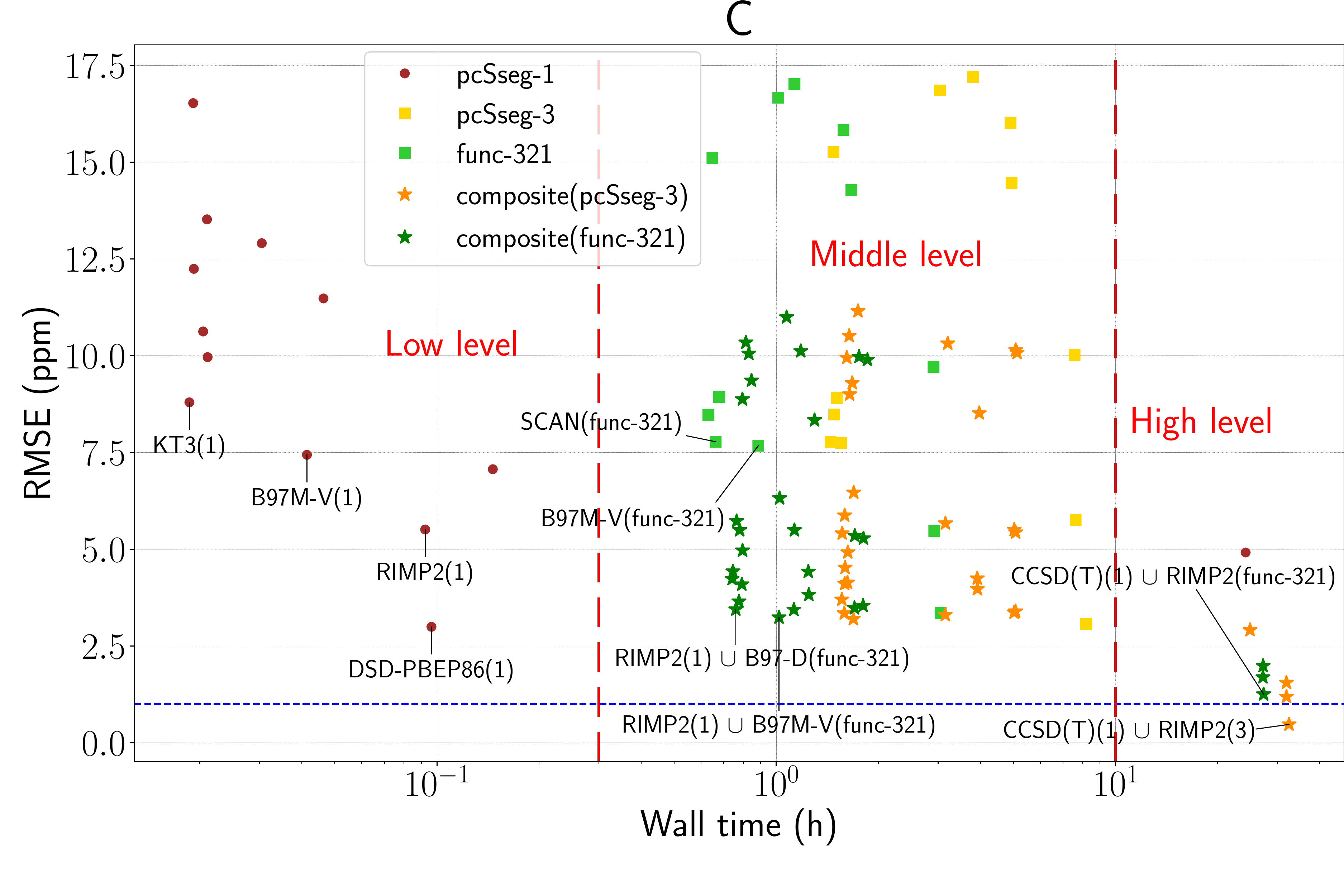}
    \caption{Comparison of the RMSEs for carbon shieldings (in ppm) of different methods with different basis sets relative to CCSD(T)/pcSseg-3 on the NS372 set against the average wall time (in hours) for molecules with eight non-hydrogen atoms calculated with four physical cores. ``(X)" in the method labels represents the basis set pcSseg-X. Only the recommended methods are labeled here. The blue horizontal lines indicate the target error: 1 ppm for C nuclei.}
    \label{fig:overall_C}
\end{figure}
\begin{figure}
    \centering
    \includegraphics[width=0.7\textwidth]{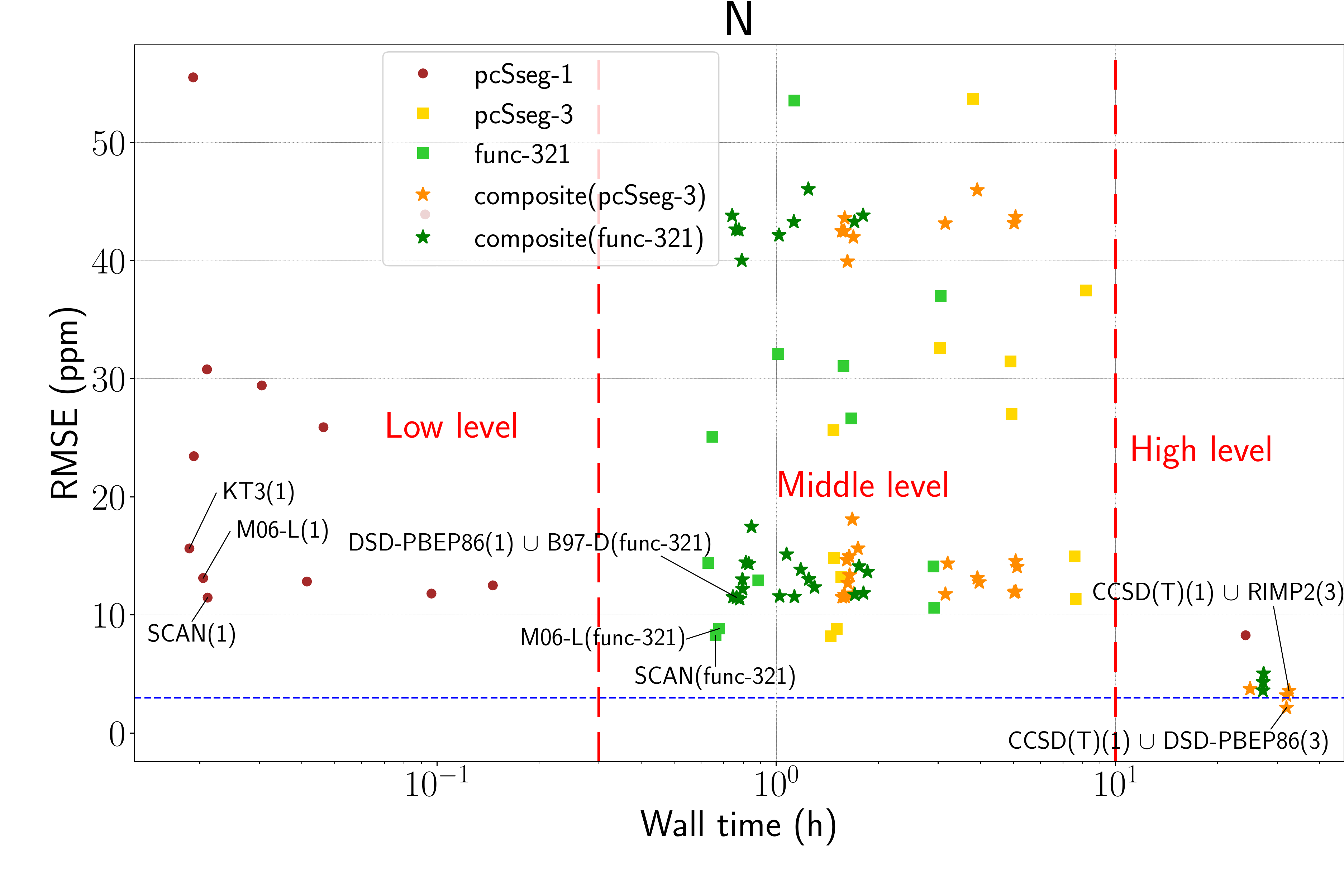}
    \caption{Comparison of the RMSEs for nitrogen shieldings (in ppm) of different methods with different basis sets relative to CCSD(T)/pcSseg-3 on the NS372 set against the average wall time (in hours) for molecules with eight non-hydrogen atoms calculated with four physical cores. ``(X)" in the method labels represents the basis set pcSseg-X. Only the recommended methods are labeled here. The blue horizontal lines indicate the target error: 3 ppm for N nuclei.}
    \label{fig:overall_N}
\end{figure}
\begin{figure}
    \centering
    \includegraphics[width=0.7\textwidth]{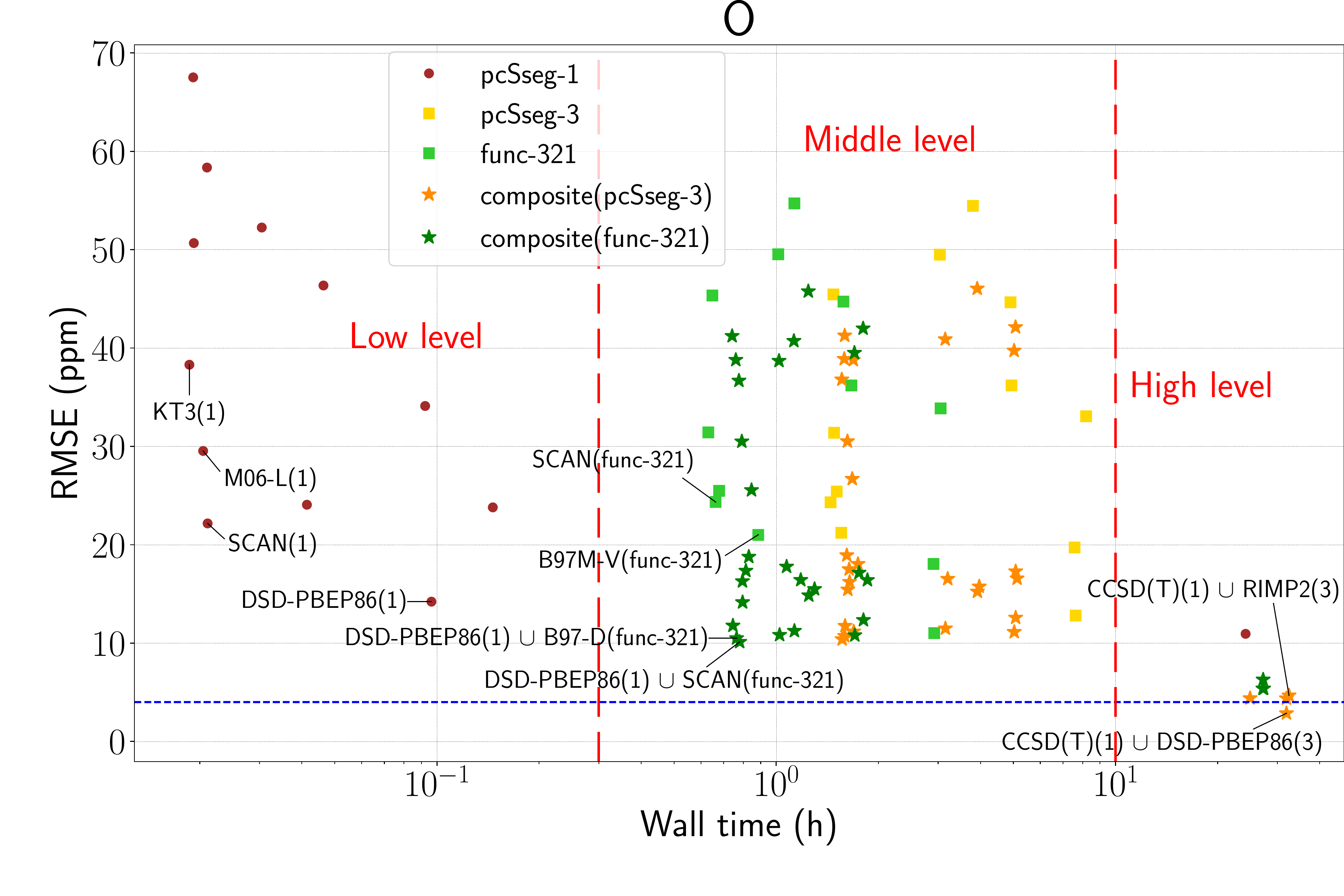}
    \caption{Comparison of the RMSEs for oxygen shieldings (in ppm) of different methods with different basis sets relative to CCSD(T)/pcSseg-3 on the NS372 set against the average wall time (in hours) for molecules with eight non-hydrogen atoms calculated with four physical cores. ``(X)" in the method labels represents the basis set pcSseg-X. Only the recommended methods are labeled here. The blue horizontal lines indicate the target error: 4 ppm for O nuclei.}
    \label{fig:overall_O}
\end{figure}

Figures~\ref{fig:overall_H}, \ref{fig:overall_C}, \ref{fig:overall_N}, and \ref{fig:overall_O} show the RMSEs of all tested methods across the hydrogen, carbon, oxygen, and nitrogen nuclei respectively against their average wall time for molecules with 8 non-hydrogen atoms using MPI parallelization on four physical cores. These methods can be divided into three levels according to their computational costs, and the best methods for each of the three levels are also in order of overall accuracy. We only labeled the recommended methods for each level here and full numerical data is contained in Tables~S1.3 and S1.4.

Starting with the low-level methods, for protons (H), PBE0(1) is preferred, while DSD-PBEP86(1) is the best or near-best functional on all other nuclei.  However, the accuracy of the low-level methods is not acceptable for practical use and the result of these methods can only be used for rough calculations or possibly as input for machine learning networks. If DHDFT (DSD-PBEP86) is not cheap enough for these purposes, then semi-local functionals (like KT3, B97M-V, SCAN, and M06-L) are recommended.

For the middle-level methods, it is clear that LDBS (green points) can significantly reduce calculation time compared with the pcSseg-3 basis set (orange/yellow points) while maintaining nearly the same accuracy. Except for N nuclei, the composite methods corrected by RIMP2 or DHDFT (star points) can decrease the RMSEs a lot compared with the original methods (some square points) while increasing compute costs only a little. We have labeled the recommended methods for each nucleus in Figures~\ref{fig:overall_H}, \ref{fig:overall_C}, \ref{fig:overall_N}, and \ref{fig:overall_O}. If all the four kinds of nuclei are wanted in the lowest computational time, we suggest RIMP2(1) $\cup$ B97-D(func-321) for the C nucleus and DSD-PBEP86(1) $\cup$ B97-D(func-321) for other nuclei. It is reasonable to use both RIMP2 and DSD-PBEP86 as $T_{\text{high}}$ because their cost is small compared with that of B97-D(func-321) for reasonable molecule sizes. Only proton shieldings achieve their target accuracy with the recommended middle-level methods.

Regarding high-level methods, we note that CCSD(T)(1) $\cup$ RIMP2(3) performs the best on H and C nuclei and CCSD(T)(1) $\cup$ DSD-PBEP86(3) performs the best on N and O nuclei. They are also the only methods that reach the target accuracy for C, N, and O nuclei. When the molecule of interest has more than 4 non-hydrogen atoms, the CCSD(T)(1) part of the composite method will be more expensive than the RIMP2(3) or DSD-PBEP86(3) part (Table~S1.4). Therefore, we can use the two composite methods simultaneously. If predicting all types of nuclei by one method is needed, researchers can use CCSD(T)(1) $\cup$ RIMP2(3) since it is also close to the target error of N and O nuclei. It is easy to understand understand why CCSD(T)(1)$\cup$RIMP2(3) performs well on H and C nuclei given that MP2 and CCSD(T) have comparable basis set convergence trends (as described in Section 3.1) and the basis set correction of MP2 becomes more appropriate as a result. The relative poor performance of CCSD(T)(1)$\cup$RIMP2(3) on N and O nuclei can be possibly ascribed to the relatively poor performance of MP2 itself (the RMSE of MP2 is nearly three times that of DSD-PBEP86). Additionally, the LDBS technique only marginally cuts down on time but doubles or even triples the errors. Therefore, with the codes used here, the LDBS models that we have tested cannot be recommended for high-level methods.

A comparison of RMSEs of chemical shifts has been made using different molecules as reference in Table~S1.5 for interested readers. Due to partial error cancellation, the majority of the low-level methods can obtain a reduced RMSE, and the ranking of middle- and low-level methods is affected. Generally, the change for N and O nuclei is smaller than that for H and C nuclei because the larger paramagnetic terms for N and O lead to a less pronounced error cancellation effect. If one wants to calculate chemical shifts more accurately by a middle- or low-level method, we recommend choosing a small reference molecule whose nuclei are in a similar electronic environment to the target system and choosing one best method using this reference in Table~S1.5. However, our recommendations for high-level methods remain unchanged, because these methods are already very accurate and the influence of error cancellation is minimal.

\section{Conclusions} \label{sec:conclusion}

Building on the work of prior researchers on locally density basis sets (LDBS) and composite methods for NMR shielding calculations, we have investigated the most effective strategies to use under various time and accuracy requirements. Regarding basis sets and LDBS approaches, our main conclusions are as follows:
\begin{enumerate}
    \item We demonstrated that there is relatively little difference in either accuracy or compute cost between the pcSseg-n and pcS-n basis sets for n $\geq 1$. Simply because pcSseg-n is about 5-10\% cheaper than pcS-n, the pcSseg-n series was selected for this work.
    \item We assessed three different LDBS partition schemes and concluded (pcSseg-)func-321 preferable, which allocates pcSseg-3, pcSseg-2, and pcSseg-1 basis sets to the target group, nearest-neighbor groups, and more remote groups respectively after splitting the molecule by functional groups.
\end{enumerate}

Our main results were calculations on a large set of NMR shieldings on H, C, N, and O nuclei to evaluate the compute costs and accuracy of many methods employing the LDBS and composite methods. We divided the methods into three levels:
\begin{enumerate}
    \item To reach the desired high accuracy (0.1 ppm for H, 1 ppm for C, 3 ppm for N, and 4 ppm for O), we recommend CCSD(T)(1) $\cup$ RIMP2(3) for H and C, and CCSD(T)(1) $\cup$ DSD-PBEP86(3) for N and O nuclei.
    \item At a middle level of cost and accuracy, we recommend RIMP2(1) $\cup$ B97-D(func-321) for C and  DSD-PBEP86(1) $\cup$ B97-D(func-321) for other nuclei. This reaches a high level of accuracy for H and is 2-3 times larger than the target for C, N, and O.
    \item  When the lowest compute cost is essential, such as to generate a large data set of chemical shielding constants, the best option appears to be the use of a  semi-local functional (like KT3, B97M-V, SCAN, and M06-L) with the pcSseg-1 basis set.
\end{enumerate}

In terms of caveats, this work only involves elements lighter than Ar. Therefore it is an open question how well our results will transfer to molecules containing heavier elements, and in particular transition metal atoms, since they can have more challenging electronic structure. For example, CCSD(T)/pcSseg-3 might not be a sufficiently accurate reference for transition-metal oxo complexes which show large static correlation effects. Relativistic effects are also not taken into account in our work and will make some contribution for heavier elements, particularly when the shielding is evaluated at that nucleus.

We note that all conclusions drawn in this work are dictated by the performance characteristics of the codes used to evaluate the shieldings. Advances in those codes, or the development of new algorithms, could significantly change some of our recommendations. It also seems clear that  the development of new electronic structure methods which offer improved trade-offs between cost and accuracy would be highly desirable to further advance calculations whose cost is at the low or middle levels.

\section*{Supporting Information}

Additional information and figures (SI.pdf)

S1-data\_analysis.xlsx

S2-raw\_data.xlsx

\begin{acknowledgement}
This work was primarily supported by funding from the National Institute of General Medical Sciences (National Institutes of Health) under grant number 5U01GM121667]. Additional support to complete the project came from the Director, Office of Science, Office of Basic Energy Sciences, of the U.S. Department of Energy through the Gas Phase Chemical Physics Program, under Contract No. DE-AC02-05CH11231. This research used computational resources of the National Energy Research Scientific Computing Center, a DOE Office of Science User Facility supported by the Office of Science of the U.S. Department of Energy under Contract No. DE-AC02-05CH11231.
\end{acknowledgement}

\clearpage

\bibliography{a-main}
\end{document}